\documentclass{aa}  

\usepackage{graphicx}
\usepackage{txfonts,textcomp}
\usepackage{natbib}

\usepackage{multirow}
\makeatletter

\newcommand{\Rmnum}[1]{\expandafter\@slowromancap\romannumeral #1@}
\makeatother

\begin{document}

   \title{Inside-out growth and the kiloparsec-scale star formation main sequence for low-surface-brightness disk galaxies in MaNGA}
   \titlerunning{disk LSBGs in MaNGA}

   \author{Bing-qing Zhang \inst{1} \email{bqzhang@bao.ac.cn}
   \and Ze-hao Zhong \inst{2} \email{zhzhong@nao.cas.cn}
        \and Yi-nan Zhu \inst{1} \email{astro\_yz@163.com}
        \and Hong Wu \inst{1} \email{hwu@bao.ac.cn}
        }

   \institute{Key Laboratory of Optical Astronomy, National Astronomical Observatories, Chinese Academy of Sciences, 20A Datun Road, Chaoyang District, Beijing, 100101, China
   \and School of Astronomy and Space Science, University of Chinese Academy of Sciences, Beijing, 100049, China}

   \date{Received April xx, 2026}

  \abstract  
   {Low surface brightness galaxies (LSBGs) are very faint objects in the universe whose formation and evolution have long remained an unresolved problem. However, so far the spatially resolved properties and formation mechanisms of LSBGs have been little studied. Only a few works have studied the spatially resolved properties of single LSBGs and there is a lack of statistical analyses for a LSBG sample.}
   {This paper aims to construct a LSBG sample based on integral field unit (IFU) data, to investigate the spatial distribution of galaxy properties, and to explore the formation mode and evolutionary process of different galaxy regions for LSBGs.}
   {We selected a late-type, face-on, and disk-dominated galaxy sample from the MaNGA survey, composed of 38 LSBGs and 216 high surface brightness galaxies (HSBGs) by adopting a central surface brightness threshold $\mu_{0}(g) = {\rm 22.0\ mag\ arcsec^{-2}}$. Based on maps derived from pipe3D products of star-forming areas in galaxies, we investigated the global and resolved star formation main sequence (SFMS and rSFMS) and radial profiles of the stellar mass density ($\mathrm{\Sigma_{*}}$), star formation rate density ($\mathrm{\Sigma_{SFR}}$), and specific star formation rate ($\mathrm{\Sigma_{sSFR}}$).}
   {The radial profiles of $\mathrm{\Sigma_{*}}$ and $\mathrm{\Sigma_{SFR}}$ present negative gradients, while those of $\mathrm{\Sigma_{sSFR}}$ present positive gradients. Compared to HSBGs, the LSBGs show similar, flatter, and steeper slopes in $\mathrm{\Sigma_{*}}$, $\mathrm{\Sigma_{SFR}}$, and $\mathrm{\Sigma_{sSFR}}$ gradients, respectively. The very flat $\mathrm{\Sigma_{SFR}}$ gradients of LSBGs, whose slopes range from -0.1 to -0.2, indicates uniform star formation throughout the whole galaxy, and the positive $\mathrm{\Sigma_{sSFR}}$ gradients reflects their ``inside-out'' growth.
   Our LSBGs and HSBGs follow the same global and resolved SFMS as the full sample, and central surface brightness shows no influence on both SFMS and rSFMS. For our full sample and two subsamples (HSBGs and LSBGs), the slopes of global SFMS and rSFMS are consistent, which indicates that the SFMS still hold on a kiloparsec scale and the star formation activities are regulated by local events. The treatment of different excitation sources in galaxies affects the slope of the SFMS and rSFMS, and using star-forming areas only is more advisable.}
   {The disk LSBGs in our sample have experienced an ``inside-out'' formation process. At the current stage, their star formation activities are at a very low level, and they maintain a remarkably flat star formation rate (SFR) gradient. Our LSBGs and HSBGs follow the same SFMS, which also holds locally, indicating that the evolution of our disk LSBGs is regulated by local properties, while stellar mass and central surface brightness have a negligible impact on the SFMS.}


   \maketitle
   \nolinenumbers

\section{Introduction}
Low surface brightness galaxies (LSBGs) are faint objects in the universe whose central surface brightness of the disk component is at least one magnitude fainter than the sky brightness \citep{1997ARAA...35..267I}; nevertheless, they play an important role in understanding galaxy formation and evolution. LSBGs constitute most of the total number density of low-mass galaxies \citep{1997AJ....114..635D, 2000ApJ...529..811O} and around 10\% of the cosmic baryon budget \citep{2004MNRAS.355.1303M}. Research into the global properties of LSBGs has revealed that they hold low star formation activity \citep{1993AJ....106..548V, 2011ApJ...728...74G, 2018ApJS..235...18L, 2019ApJS..242...11L}, low gas metallicity \citep{1994AJ....107..530M, 2010MNRAS.409..213L}, rich neutral hydrogen (H\Rmnum{1}) gas \citep{1996MNRAS.283...18D, 2001AJ....122.2318B, 2004AJ....128.2080O, 2015AJ....149..199D}, and few CO molecules \citep{2003ApJ...588..230O, 2018MNRAS.476.4488H}. 

Among galaxy properties, stellar mass ($\mathrm{M_{*}}$) is an important parameter because many other parameters are related to it, which provides some famous scale relations. The most intuitive process that stars are formed directly from molecular clouds provides the Kennicutt-Schmidt law (KS), the relation between star formation rate (SFR) and molecular gas mass ($\mathrm{M_{gas}}$), and the molecular gas main sequence (MGMS), the relation between $\mathrm{M_{gas}}$ and $\mathrm{M_{*}}$. The progress of stellar mass accumulation and chemical enrichment establishes connection between gas metallicity and $\mathrm{M_{*}}$ (MZR). The star formation main sequence (SFMS) is the relation between SFR and $\mathrm{M_{*}}$, although the SFMS may not be the most fundamental relation, instead of a secondary relation between KS law and MGMS \citep{2019ApJ...884L..33L, 2021MNRAS.501.4777E, 2021MNRAS.503.1615S, 2022MNRAS.510.3622B, 2023MNRAS.518.4767B, 2026Univ...12...60G}, it is also very valuable for galaxies that cannot get reliable molecular detections.

However, a question remains on whether the local properties of a galaxy resemble its global characteristics, and how global and local properties regulate galaxy evolution. Integral field unit (IFU) spectroscopic observations enable spatially resolved analyses of galaxy properties. Some researchers find many global scaling relations present resolved counterparts, the global as a consequence of the local one. For each of the SFMS, KS law, and MGMS, the slopes, intercepts, and scatters remain similar across the different scales explored \citep{2020ARA&A..58...99S, 2021MNRAS.503.1615S, 2021AA...650A.134P}. However, \citet{2019MNRAS.488.3929C} find that for star-forming areas (SFAs), which are regions in a galaxy where ionization is dominated by young stars, there is a trend to lower values of SFR with earlier morphological types, with morphology partially influencing the slope and intercept of rSFMS. \citet{2023MNRAS.519.1149B} find the global MZR does not simply stem out of its resolved counterparts, and that global properties and processes likely also contribute to the local metallicity, in addition to local production and retention. As for LSBGs, the dwarf LSBGs in \citet{2017ApJ...851...22M} and the disk component of giant LSBGs in \citet{2023ApJ...959..105D} follows the SFMS defined by high surface brightness galaxies (HSBGs), but the gas-rich LSBGs in \citet{2025PASP..137j4102C} do not. Up to now, only a few studies have investigated the global relations for LSBGs, and there is a lack in resolved relations of LSBGs. 

The IFU observations also enable analyses for radial distributions of galaxy properties. The radial distributions of galaxy properties (e.g., stellar mass ($\mathrm{M_{*}}$), stellar age ($\mathrm{A_{*}}$), stellar metallicity ($\mathrm{Z_{*}}$), gas-phase oxygen abundance (12+log(O/H)), and SFR) serve as powerful diagnostics of the evolutionary process of galaxies. The negative gradient of the gas-phase oxygen abundance in disk star formation galaxies (SFGs) is direct evidence of an inside-out formation of these galaxies \citep{1989MNRAS.239..885M, 1999MNRAS.307..857B}. In addition, studies show that massive galaxies ($\mathrm{\geqslant 10^{9.5 \sim 10} M_{\odot}}$) follow an inside-out growth \citep{2005ApJ...635..959B, 2013ApJ...771L..35V}, whereas lower-mass galaxies ($\mathrm{\leqslant 10^{8 \sim 9} M_{\odot}}$) present a possible transition from inside-out toward outside-in \citep{2013ApJ...764L...1P}. As the stellar mass of LSBGs spans a large range, approximately $\mathrm{10^{7 \sim 10.5} M_{\odot}}$ \citep{2020AJ....159..138D, 2024RAA....24a5018Z}, there may exist different types of growth mode among LSBGs. Thus, it is necessary to investigate the radial profiles of LSBGs in order to understand their formation mechanisms and evolutionary pathways.

In this paper, we aim to check the global and resolved SFMS, as well as radial distributions of stellar mass density ($\mathrm{\Sigma_{*}}$), SFR density ($\mathrm{\Sigma_{SFR}}$), and the specific SFR ($\mathrm{\Sigma_{sSFR}}$) of LSBGs. The rest of this paper is arranged as follows. Section \ref{sec:Data and Sample} describes our LSBGs and control sample (HSBGs) selection. Section \ref{sec:Mesurement of Parameters} describes calculation procedures of $\mathrm{\Sigma_{SFR}}$ and $\mathrm{\Sigma_{sSFR}}$. In Section \ref{subsec:radial profile}, we show radial distributions of $\mathrm{\Sigma_{*}}$, $\mathrm{\Sigma_{SFR}}$, and $\mathrm{\Sigma_{sSFR}}$, and in Section \ref{subsec:SFMS}, we show the results of global and resolved SFMS. In Section \ref{sec:Discussion}, we discuss the effects of different excitations within galaxies on the SFMS and the limitation of our LSBG sample. Finally, we summarize this paper in Section \ref{sec:Summary}. The Hubble time adopted in this paper is 13.7 Gyr.

\section{Data and sample} \label{sec:Data and Sample}
\subsection{The MaNGA survey} \label{subsec:The MaNGA Survey}
The Mapping Nearby Galaxies at Apache Point Observatory (MaNGA) survey involves IFU spectroscopic observations of over 10,000 nearby galaxies \citep{2015ApJ...798....7B}. It achieves a spatial resolution of $2.5^{\prime\prime}$, which corresponds to physical scales ranging from 0.5 kpc (at $z = 0.01$) to 5 kpc (at $z = 0.1$). The spectral coverage spans 3600 $\sim$ 10300 $\mathring{A}$, with a signal-to-noise ratio (S/N) reaching 20 at the red end and 36 at the blue end.

\citet{2022MNRAS.509.4024D} provides two value added catalogs (VACs). One is the MaNGA PyMorph photometric VAC (MPP-VAC-DR17), which performs S\'{e}rsic and S\'{e}rsic+exponential fits to the images of galaxies in MaNGA DR17, providing very useful parameters to calculate the central surface brightness ($\mu_{0}$) of a galaxy, such as total flux, half-light radii ($\rm{R_{e}}$), axis-ratio (b/a), and position angle (PA). Another is the MaNGA Deep Learning Morphological VAC (MDLM-VAC-DR17), which performs morphological classification for the same galaxies by using a deep learning method, helping to distinguish between late-type and early-type galaxies, edge-on and face-on galaxies, and barred and unbarred galaxies.

\subsection{Samples: LSBG and HSBG selection} \label{subsec:Sample Selction}
In the MPP-VAC-DR17 catalog, there is a useful parameter, ``FLAG\_FIT,'' to represent the preference of the two models (the single S\'{e}rsic model and the S\'{e}rsic+exponential model). FLAG\_FIT = 1 indicates that the parameters from the single S\'{e}rsic fit are preferred, FLAG\_FIT = 2 indicates that a galaxy is clearly made of two components (S\'{e}rsic+exponential model is preferred), FLAG\_FIT = 0 indicates that the two models are both acceptable, and FLAG\_FIT = 3 indicates that both models have failed. 

In this paper, we focus on disk-dominated LSBGs because previous studies have shown that LSBGs are mostly disk-dominated galaxies, which can be well fitted by an exponential profile \citep{1994AJ....107..530M, 1995MNRAS.274..235D, 1997AJ....113.1212O, 2015AJ....149..199D, 2024RAA....24a5018Z}. In recent years, LSBGs with bulges have also attracted attention \citep{2018MNRAS.478.4657P, 2018ApJ...857..104G, 2019MNRAS.485..796M, 2021ApJS..252...18T, 2024RAA....24e5015D}, and we also plan to select such galaxies from MaNGA in the future for comparison with disk-dominated LSBGs. However, this is beyond the scope of the present work. Since our goal is to study disk-dominated LSBGs, galaxies that are well fitted by a single S\'{e}rsic model are more suitable. The sample selection procedures are described in the following.

Firstly, from all of the MaNGA galaxies, we selected galaxies that are labeled as $\rm FLAG\_FIT = 1\ or\ 0$ in MPP-VAC-DR17 catalog, which means that in their case ``the single S\'{e}rsic model is better than (or equal to) the S\'{e}rsic+exponential model'. This resulted a subsample containing 6815 galaxies. 

Secondly, we adopted the following criteria to make further selections from the sample in the previous step:
\begin{enumerate}
        \item Select late-type galaxies. In this step, we used the parameters in the MDLM-VAC-DR17 catalog: T-Type $\rm >$ 0, $\rm Visual\_Class$ = 3, and $\rm P_{LTG} \geqslant 0.7$.
        \item Select face-on galaxies. $\rm P_{edge-on} \leqslant 0.2$ (in MDLM-VAC-DR17 catalog) and axis ratio of S\'{e}rsic model $b/a \geqslant 0.5$ (in MPP-VAC-DR17 catalog).
        \item Select disk galaxies according to S\'{e}rsic n index $\rm 0.5 \leqslant n \leqslant 1.5$.
        \item The observed field of view (FoV) should be large enough to cover most of the galaxy area, but not so large that many outer fibers are useless: $\rm 2R_{e} \leqslant FoV \leqslant 5R_{e}$. We only used observations taken with IFUs composed of 91 and 127 fibers to obtain as many spatial sampling points as possible.
        \item Check the excitation source of the galaxy center to exclude galaxies with potential active galactic nuclei (AGN). In this step, we draw a Baldwin-Phillips-Terlevich (BPT) diagram by using the fluxes of four emission lines ([N\Rmnum{2}], H$\alpha$, [O\Rmnum{3}], and H$\beta$) of a fixed 2.5$^{\prime\prime}$ diameter aperture in the galaxy center. We exclude galaxies with locations above the pure star formation curve, $\rm log([O\Rmnum{3}]/H\beta) = 1.3 + 0.61/(log([N\Rmnum{2}]/H\alpha) - 0.05)$ \citep{2003MNRAS.346.1055K}.
\end{enumerate}
After applying the five selection steps sequentially, the numbers of the remaining galaxies are 3165, 1886, 779, 591, and 438, respectively. The fractions of removed galaxies at each step relative to the initial sample (6815) are 53.6\%, 18.8\%, 16.2\%, 2.8\%, and 2.2\%, respectively. The remaining 438 galaxies are late-type, face-on, disk-dominated, and star-forming galaxies.

Thirdly, we selected the LSBG and HSBG samples using the following methods:
\begin{enumerate}
        \item LSBG sample. We set a limit on the central surface brightness of $\mu_{0}{\rm (g) \geqslant 22.0\ mag\ arcsec^{-2}}$ to select LSBGs. The morphological parameters needed for $\mu_{0}$ calculation are from the MPP-VAC-DR17 catalog, and the method is presented in our previous work (Section 2.3 of \citet{2024RAA....24a5018Z}). We obtained 53 LSBGs, and the stellar mass ($\rm log(M_{*}/M_{\odot})$) of our LSBG sample spans from 8.93 to 10.63, with $96.2\%$ (51/53) of them having masses between 9 and 10.5, and only two LSBGs beyond this range. In the following sections, we group the galaxies by mass and for convenience we removed these two galaxies from our sample. Consequently, we obtained a sample with 51 LSBGs, whose stellar mass spans from $10^{9}$ to $10^{10.5}$ $\rm M_{\odot}$.
        \item HSBG sample. Of the 438 galaxies obtained from the second step, we selected HSBGs by applying the criteria $\mu_{0}{\rm (g) < 22.0\ mag\ arcsec^{-2}}$ and $\rm 9 \leqslant log(M_{*}/M_{\odot}) \leqslant 10.5$. We got 275 HSBGs.
\end{enumerate}

\begin{figure}
        \centering
        \includegraphics[width=7cm]{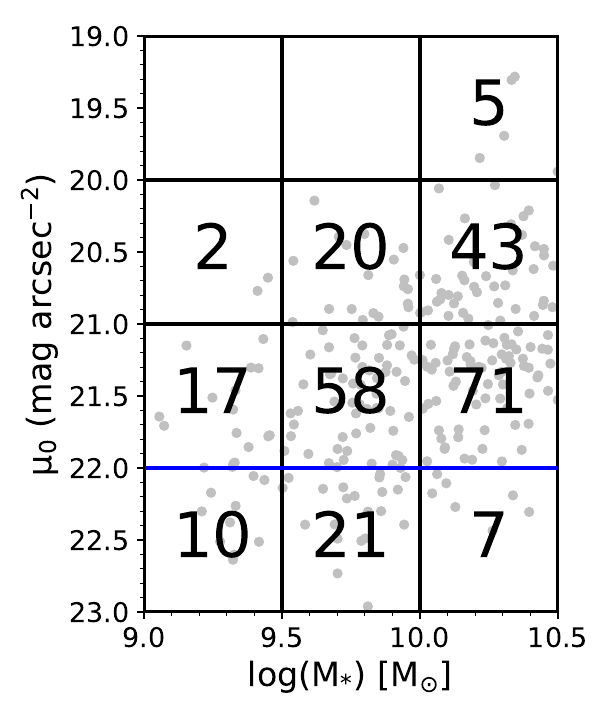}
        \caption{Number of galaxies in each $\mathrm{M_{*}}$ - $\mu_{0}$ bin. The blue line, $\mu_{0} = \mathrm{22.0\ mag\ arcsec^{-2}}$, represents the central surface brightness threshold that separates LSBGs and HSBGs.}
        \label{fig:group_number}
\end{figure}

Finally, we performed a visual inspection of FoV images of the 51 LSBGs and 275 HSBGs. We excluded galaxies with significant asymmetries in their morphology and overlaid by other galaxies or stars that cannot be effectively masked. As a result, our final LSBG sample contains 38 galaxies and the HSBG sample contains 216 galaxies. Some of the parameters of our LSBG sample are shown in Table \ref{tab:LSBG_info}.
Additionally, our samples are divided into several groups according to their total stellar mass ($\mathrm{M_{*}}$) and central surface brightness ($\mu_{0}$). We list the number of galaxies in each $\mathrm{M_{*}}$ - $\mu_{0}$ bin in Figure \ref{fig:group_number}.

\section{Mesurement of parameters} \label{sec:Mesurement of Parameters}

\subsection{Star formation rate density, $\mathrm{\Sigma_{SFR}}$} \label{subsec:Sigma_SFR calculation}
The H$\alpha$ emission line is one of the most reliable tracers of the current SFR for H\Rmnum{2} regions. For faint galaxies, especially LSBGs, an accurate measurement of SFR is important. Here, spaxels whose $\rm S/N \geqslant 3$ for H$\alpha$ and H$\beta$ were available. We calculated the SFR using the following procedure.

Firstly, we estimated the internal dust attenuation by using the ``Balmer decrement'' method:
\begin{equation}
        \rm
        E(B-V) = \frac{2.5}{R_{H\beta} - R_{H\alpha}} \times log\bigg[ \frac{(H\alpha/H\beta)_{obs}}{(H\alpha/H\beta)_{intr}}\bigg]
\end{equation}

\begin{equation}
        \rm
        A_{H\alpha} = R_{H\alpha} \times E(B-V)
,\end{equation}
where the intrinsic $\mathrm{H\alpha/H\beta}$ ratio is 2.87, $\mathrm{R_{H\alpha} = 2.468,}$ and $\mathrm{R_{H\beta} - R_{H\alpha} = 1.163}$ \citep{2001PASP..113.1449C}. The dust-corrected H$\alpha$ luminosity is
\begin{equation}
        \rm
        L_{H\alpha\_corr} = L_{H\alpha\_obs} / 10^{-0.4 A_{H\alpha}}
.\end{equation}

Then, we calculated the SFR by using the dust-corrected H$\alpha$ flux \citep{1998ARA&A..36..189K}:
\begin{equation}
        \rm
        SFR(M_{\odot}\ yr^{-1}) = 7.9 \times 10^{-42} L_{H\alpha\_corr} (erg\ s^{-1})
.\end{equation}

Finally, we divided the SFR by the area of one spaxel to get the $\mathrm{\Sigma_{SFR}}$:
\begin{equation}
        \rm
        \Sigma_{SFR}(M_{\odot}\ yr^{-1}\ kpc^{-2}) = \frac{SFR(M_{\odot}\ yr^{-1})}{S(kpc^{2})}
.\end{equation}

\subsection{Specific star formation rate, $\mathrm{\Sigma_{sSFR}}$} \label{subsec:Sigma_sSFR calculation}
The stellar mass density ($\mathrm{\Sigma_{*}}$) map used in this paper was obtained from stellar population analysis results of the data products of the pipe3D team, and is corrected for extinction by dust; see Section 5.1.1 in \citet{2022ApJS..262...36S} for details. The length of each spaxel is 0.5$^{\prime\prime}$; the area of one spaxel in units of $\mathrm{kpc^{2}}$  is calculated according to galaxy distance obtained from the MaNGA Data Analysis Pipeline (DAP) catalog \citep{2019AJ....158..231W}. Thus, the unit of $\mathrm{\Sigma_{*}}$ in this paper is $\mathrm{M_{\odot}\ kpc^{-2}}$. 

The $\mathrm{\Sigma_{sSFR}}$ is the ratio of $\mathrm{\Sigma_{SFR}}$ to $\mathrm{\Sigma_{*}}$. This quantity represents the reciprocal of the time required for the galaxy to form its current stellar mass at the present SFR:
\begin{equation}
        \rm
        \Sigma_{sSFR}(yr^{-1}) = \frac{\Sigma_{SFR}(M_{\odot}\ yr^{-1}\ kpc^{-2})}{\Sigma_{*}(M_{\odot}\ kpc^{-2})}
.\end{equation}

\section{Results and analyses} \label{sec:Results}
\subsection{Radial profiles} \label{subsec:radial profile}

\begin{figure*}[htbp]
        \includegraphics[width=\hsize]{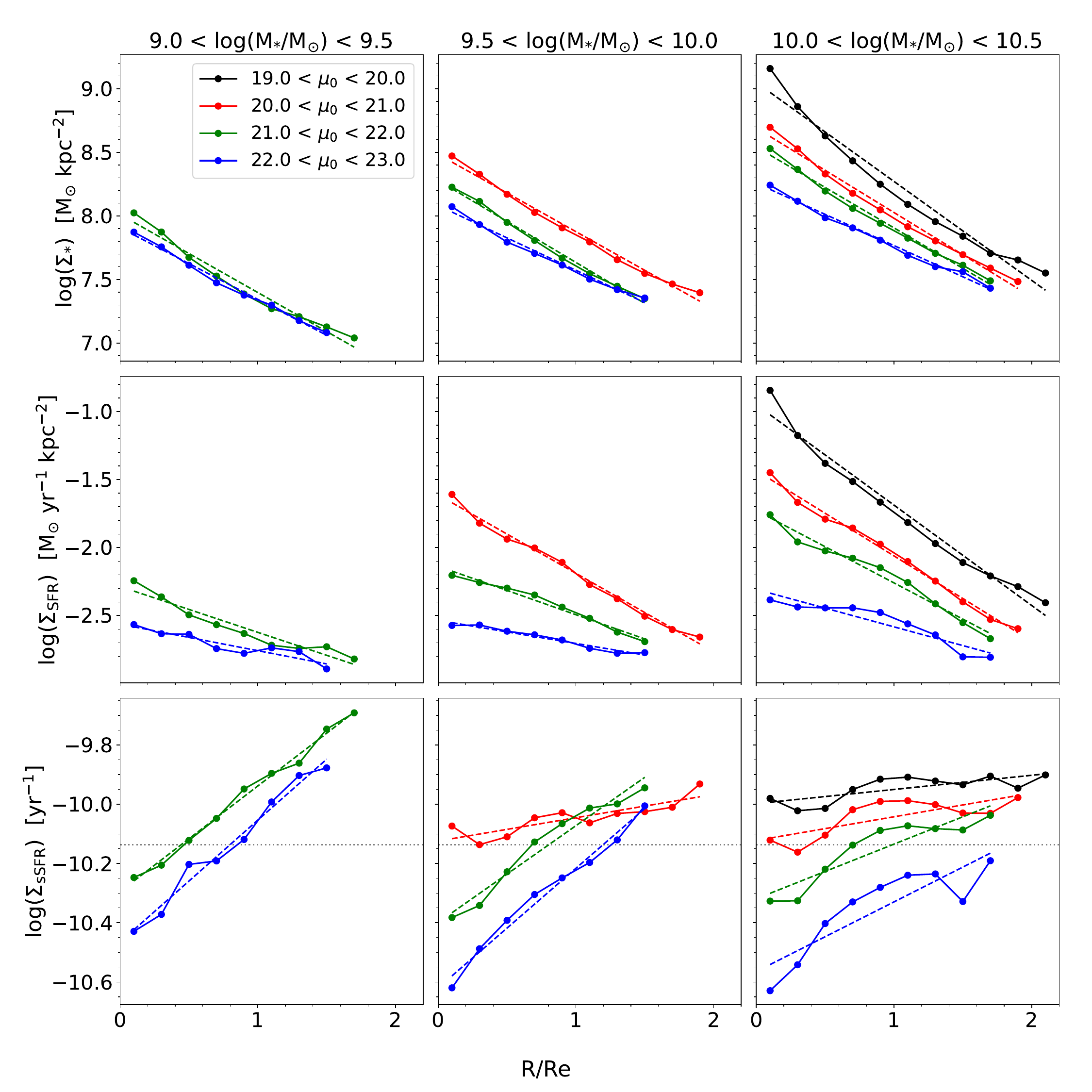}
        \caption{Average radial profiles in each $\mathrm{M_{*}}$ - $\mu_{0}$ bin of stellar mass density ($\mathrm{\Sigma_{*}}$; the first row), star formation rate density ($\mathrm{\Sigma_{SFR}}$; the second row), and specific star formation rate ($\mathrm{\Sigma_{sSFR}}$; the third row). The three columns cover different stellar mass ranges of galaxies, and the colors of the lines represent different central surface brightnesses ($\mu_{0}$). The dashed lines are the linear fitting of the radial profiles, and the fitting parameters are listed in Table \ref{tab:radial linear fitting}. The dotted gray line in each panel of the third row is the characteristic $\mathrm{\Sigma_{sSFR}}$, -10.137, which means galaxies can form their current stellar mass with their current SFR throughout Hubble time.}
        \label{fig:radial}
\end{figure*}

In this section, we check the radial distributions of stellar mass density ($\mathrm{\Sigma_{*}}$), star formation rate density ($\mathrm{\Sigma_{SFR}}$), and specific star formation rate ($\mathrm{\Sigma_{sSFR}}$). In the following we describe the procedures used to obtain radial profiles.

Firstly, the radial profile of each individual galaxy must be obtained. We adopted a series of elliptical apertures centered by the galaxy center, with 0.2 $\mathrm{R/R_{e}}$ intervals, to get the average values in each of the elliptical annuli. In this step, the normalized radius ($\mathrm{R/R_{e}}$) of each spaxel from the galaxy center was provided by the MaNGA DAP, where $\mathrm{R_{e}}$ corresponds to Petrosian $r$-band R50 from the NASA Sloan Atlas (NSA) \citep{2019AJ....158..231W}. 

Secondly, the average radial profile of each $\mathrm{M_{*}}$ - $\mu_{0}$ bin must be obtained. For all the individual radial profiles of galaxies in the same group, we calculated their average values at every $\mathrm{R/R_{e}}$ to derive the average radial profile of that bin. In this step, we required that at each $\mathrm{R/R_{e}}$, the number of galaxies available for averaging should be greater than 80\% of the total number in the same $\mathrm{M_{*}}$ - $\mu_{0}$ bin. Note here that we excluded one group, [$\mathrm{M_{*}}$: 9.0 - 9.5; $\mu_{0}$: 20.0 - 21.0], because that group has only two galaxies and cannot support the following statistical analysis.

Finally, we performed a linear fitting for the average radial profile. The fitting parameters, including slope, intercept, root mean square error ($\sigma$), and Pearson correlation coefficient ($\rho$), are listed in Table \ref{tab:radial linear fitting}.

\begin{table*}[htbp]
        \caption{Parameters of radial linear fitting\label{tab:radial linear fitting}}
        \renewcommand{\arraystretch}{1.2}
        \centering
        \begin{tabular}{cccccccccc}
                \hline\hline
                \multicolumn{2}{c}{$\mathrm{M_{*}-\mu_{0}}$ Bins} & 
                \multicolumn{4}{c}{$\mathrm{\Sigma_{*}}$ v.s. $\mathrm{R/R_{e}}$} & 
                \multicolumn{4}{c}{$\mathrm{\Sigma_{SFR}}$ v.s. $\mathrm{R/R_{e}}$} \\
                &   & a & b & $\sigma$ & $\rho$ & a & b & $\sigma$ & $\rho$ \\      
                \hline
                \multirow{2}{*}{9.0 $\sim$ 9.5} & 21 $\sim$ 22 & -0.668$\pm$0.036 & 8.002$\pm$0.041 & 0.065 & -0.989 & -0.323$\pm$0.028 & -2.343$\pm$0.029 & 0.043 & -0.975 \\
                & 22 $\sim$ 23 & -0.602$\pm$0.029 & 7.879$\pm$0.026 & 0.037 & -0.993 & -0.168$\pm$0.034 & -2.631$\pm$0.032 & 0.044 & -0.895 \\
                \hline
                \multirow{3}{*}{9.5 $\sim$ 10.0} & 20 $\sim$ 21 & -0.620$\pm$0.016 & 8.471$\pm$0.018 & 0.028 & -0.997 & -0.540$\pm$0.018 & -1.714$\pm$0.021 & 0.032 & -0.996 \\
                & 21 $\sim$ 22 & -0.649$\pm$0.019 & 8.235$\pm$0.019 & 0.029 & -0.997 & -0.332$\pm$0.016 & -2.219$\pm$0.015 & 0.021 & -0.993 \\      
                & 22 $\sim$ 23 & -0.546$\pm$0.016 & 8.072$\pm$0.014 & 0.020 & -0.998 & -0.142$\pm$0.017 & -2.596$\pm$0.015 & 0.022 & -0.961 \\
                \hline
                \multirow{4}{*}{10.0 $\sim$ 10.5} & 19 $\sim$ 20 & -0.818$\pm$0.043 & 9.057$\pm$0.054 & 0.090 & -0.988 & -0.763$\pm$0.032 & -0.960$\pm$0.040 & 0.067 & -0.992 \\
                & 20 $\sim$ 21 & -0.668$\pm$0.020 & 8.624$\pm$0.026 & 0.042 & -0.996 & -0.558$\pm$0.017 & -1.593$\pm$0.020 & 0.032 & -0.996 \\      
                & 21 $\sim$ 22 & -0.665$\pm$0.011 & 8.512$\pm$0.013 & 0.020 & -0.999 & -0.450$\pm$0.038 & -1.896$\pm$0.039 & 0.058 & -0.976 \\      
                & 22 $\sim$ 23 & -0.547$\pm$0.016 & 8.267$\pm$0.019 & 0.029 & -0.997 & -0.253$\pm$0.044 & -2.362$\pm$0.045 & 0.068 & -0.910 \\
                \hline\hline
                \multicolumn{2}{c}{$\mathrm{M_{*}-\mu_{0}}$ Bins} & \multicolumn{4}{c}{$\mathrm{\Sigma_{sSFR}}$ v.s. $\mathrm{R/R_{e}}$} & \\
                &   & a & b & $\sigma$ & $\rho$ & & & & \\
                \hline
                \multirow{2}{*}{9.0 $\sim$ 9.5} & 21 $\sim$ 22 & 0.356$\pm$0.012 & -10.294$\pm$0.012 & 0.018 & 0.996 & & & & \\
                & 22 $\sim$ 23 & 0.411$\pm$0.026 & -10.465$\pm$0.024 & 0.034 & 0.988 & & & & \\
                \hline
                \multirow{3}{*}{9.5 $\sim$ 10.0} & 20 $\sim$ 21 & 0.079$\pm$0.018 & -10.124$\pm$0.020 & 0.032 & 0.844 & & & & \\
                & 21 $\sim$ 22 & 0.327$\pm$0.028 & -10.399$\pm$0.026 & 0.036 & 0.979 & & & & \\      
                & 22 $\sim$ 23 & 0.403$\pm$0.022 & -10.620$\pm$0.020 & 0.028 & 0.992 & & & & \\
                \hline
                \multirow{4}{*}{10.0 $\sim$ 10.5} & 19 $\sim$ 20 & 0.048$\pm$0.014 & -9.998$\pm$0.018  & 0.030 & 0.741 & & & & \\
                & 20 $\sim$ 21 & 0.080$\pm$0.024 & -10.122$\pm$0.028 & 0.044 & 0.758 & & & & \\      
                & 21 $\sim$ 22 & 0.185$\pm$0.031 & -10.319$\pm$0.032 & 0.048 & 0.913 & & & & \\      
                & 22 $\sim$ 23 & 0.235$\pm$0.050 & -10.565$\pm$0.052 & 0.077 & 0.873 & & & & \\
                \hline  
        \end{tabular}
        \tablefoot{This table lists the parameters of linear fitting ($\mathrm{y = ax + b}$) in each $\mathrm{M_{*}}$ - $\mu_{0}$ bin for stellar mass density ($\mathrm{\Sigma_{*}}$), star formation rate density ($\mathrm{\Sigma_{SFR}}$), and specific star formation rate ($\mathrm{\Sigma_{sSFR}}$). The a, b, $\sigma$, and $\rho$ represent the slope, intercept, root mean square error, and Pearson Correlation Coefficient of linear fitting, respectively. The $\sigma$ is defined as the root value of the weighted sum of squared residuals divided by degrees of freedom.}
\end{table*}

Figure \ref{fig:radial} shows the averaged radial distributions of $\mathrm{\Sigma_{*}}$, $\mathrm{\Sigma_{SFR}}$, and $\mathrm{\Sigma_{sSFR}}$, from top to bottom. The radial profiles are plotted in three columns according to their $\mathrm{M_{*}}$ and colored by $\mu_{0}$. The dashed lines are the best fit of linear fitting. The lower the $\mathrm{M_{*}}$ and $\mu_{0}$, the lower the $\mathrm{\Sigma_{*}}$ and $\mathrm{\Sigma_{SFR}}$ values throughout the whole galaxy, whereas the $\mathrm{\Sigma_{sSFR}}$ radial profiles exhibit a systematic decline with increasing $\mathrm{M_{*}}$ and decreasing $\mu_{0}$. By employing 1240 nearby star-forming galaxies from the MaNGA survey, \citet{2024ApJ...977..175L} also found the extended galaxies show lower $\mathrm{\Sigma_{*}}$, $\mathrm{\Sigma_{SFR}}$, and $\mathrm{\Sigma_{sSFR}}$ radial profiles than compact galaxies in the same mass range. The result of \citet{2024ApJ...977..175L} is consistent with ours because, at a fixed stellar mass ($\mathrm{M_{*}}$), extended galaxies with larger radii correspond to galaxies with lower $\mu_{0}$ while compact galaxies with smaller radii correspond to galaxies with higher $\mu_{0}$.

It is clear that the radial profiles of $\mathrm{\Sigma_{*}}$ and $\mathrm{\Sigma_{SFR}}$ both decrease radially. With the exception of the most massive and luminous bin, [$\mathrm{M_{*}}$: 10.0 - 10.5; $\mu_{0}$: 19.0 - 20.0], which exhibits a pronounced central concentration on $\mathrm{\Sigma_{*}}$ radial profile, other bins exhibit a characteristic exponential decline, as evidenced by their similar slopes ranging from -0.55 to -0.70. The slopes of $\mathrm{\Sigma_{*}}$ and $\mathrm{\Sigma_{SFR}}$ gradient both become steeper and the intercepts of those become larger with the increase of $\mathrm{M_{*}}$ and $\mu_{0}$, but the slope changes of the $\mathrm{\Sigma_{SFR}}$ gradient are more prominent than that of the $\mathrm{\Sigma_{*}}$ gradient. The $\mathrm{\Sigma_{SFR}}$ radial profiles of LSBGs are nearly flat, with shallow slopes ranging from -0.1 to -0.2; this indicates that their star formation activity remains uniform from the inner regions to the outskirts. \citet{2026ApJS..284...52S} also report significantly flatter gradients for LSBGs than HSBGs and uniform star formation across the disk of LSBGs. For LSBGs \citet{2013AJ....146...41S} also found a lack of correlation between H\Rmnum{2} region luminosity and distance from the galaxy center. The $\mathrm{\Sigma_{SFR}}$ is mostly flat at about $\mathrm{10^{-4}\ M_{\odot}\ yr^{-1}\ kpc^{-2}}$ for the extended disk regions of Malin 1 \citep{2024AA...681A.100J}, which also supports our conclusion, although our LSBGs are brighter (about 1 dex) than Malin 1. 

We performed a linear fitting for the $\mathrm{\Sigma_{sSFR}}$ radial profiles, although for the massive galaxies ([$\mathrm{M_{*}}$: 10.0 - 10.5]) the shape of the $\mathrm{\Sigma_{sSFR}}$ radial profiles is no longer a straight line. The nonlinear shapes do not affect the radial upward trend. The radial profiles of $\mathrm{\Sigma_{sSFR}}$ show positive gradients, the $\mathrm{\Sigma_{sSFR}}$ at the galaxy center is lower than that of their outer parts, reflecting their ``inside-out'' growth. The slopes of the $\mathrm{\Sigma_{sSFR}}$ gradient become flatter with the increase of $\mu_{0}$. The dotted gray line is the characteristic $\mathrm{\Sigma_{sSFR}}$, -10.137, which means a galaxy can form its current stellar mass with its current SFR throughout Hubble time. Galaxy regions with a $\mathrm{\Sigma_{sSFR}}$ lower than the characteristic value must have experienced either short intense or continuous low-level star formation activities in the past. However, this cannot be discerned from the radial distribution of $\mathrm{\Sigma_{sSFR}}$ alone, which only tells us that our LSBGs and HSBGs formed from the inside out. In future work, we will analyze the star formation histories (SFH) in different regions of galaxies to get a more detailed and comprehensive understanding of the formation mode of LSBGs.

\subsection{Star formation main sequence, SFMS} \label{subsec:SFMS}
\begin{figure*}
        \includegraphics[width=\hsize]{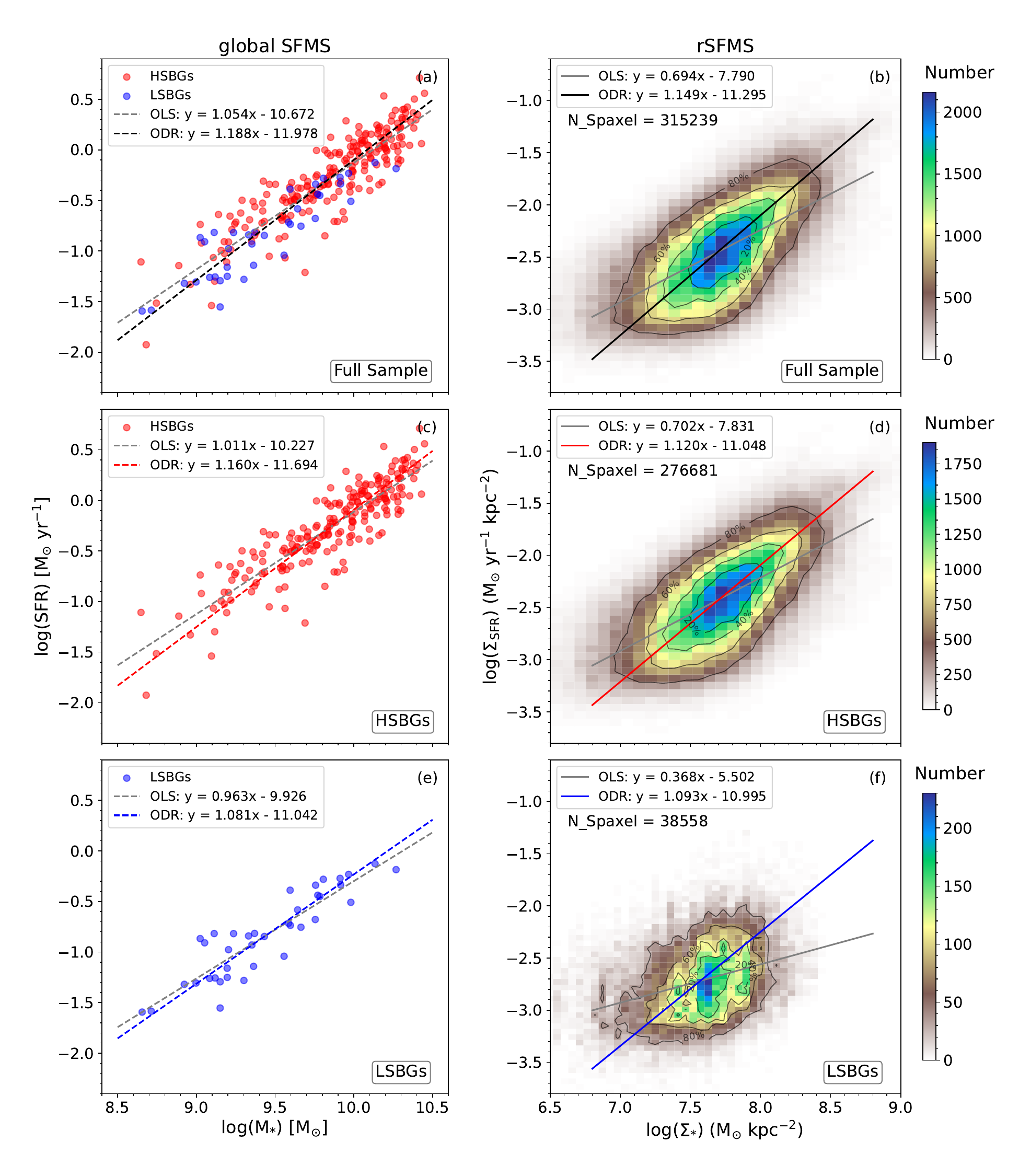}
        \caption{Global SFMS (the left column) and resolved SFMS (rSFMS; the right column) of our full sample, HSBG subsample, and LSBG subsample, from top to bottom. For the global SFMS, the red and blue scatters represent HSBGs and LSBGs, respectively. For the rSFMS, given the large amount of spaxel, 2D histograms colored by spaxel number were adopted to express the relationship. The contours enclosed 20\%, 40\%, 60\%, and 80\% of the data. We adopted two linear fitting algorithms: the OLS method and the ODR method. There are no obvious differences in the global relations between the best-fit lines of these two methods, but significant variances between these two methods are shown for the resolved relations. The ODR method is more suitable for our rSFMS fitting.}
        \label{fig:SFMS}
\end{figure*}

\begin{table*}[htbp]
        \caption{Parameters of linear fitting for SFMS\label{tab:SFMS}}
        \renewcommand{\arraystretch}{1.2}
        \centering
        \begin{tabular}{cccccccc}
                \hline\hline
                \multicolumn{2}{c}{Sample} & a & b & $\sigma$ & $\rho$ & Reference & Number of \\
                 & & & & & & & galaxies (spaxels) \\
                \hline
                \multirow{11}{*}{global}
                & Full Sample (OLS) & 1.054$\pm$0.032 & -10.672$\pm$0.310 & 0.206   & 0.902   & This work & 254 \\
                & HSBGs (OLS)       & 1.011$\pm$0.036 & -10.227$\pm$0.356 & 0.202   & 0.886   & This work & 216 \\
                & LSBGs (OLS)       & 0.963$\pm$0.079 & -9.926$\pm$0.744  & 0.187   & 0.898   & This work & 38 \\
                & Full Sample (ODR) & 1.188$\pm$0.034 & -11.978$\pm$0.332 & 0.138   &  & This work & 254 \\
                & HSBGs (ODR)       & 1.160$\pm$0.039 & -11.694$\pm$0.384 & 0.137   &  & This work & 216 \\
                & LSBGs (ODR)       & 1.081$\pm$0.084 & -11.042$\pm$0.791 & 0.131   &  & This work & 38 \\
                & SDSS DR7          & 0.76$\pm$0.01   & -7.64$\pm$0.02    &  &  & \citet{2015ApJ...801L..29R} & $\sim$ 240,000 \\
                & CALIFA            & 0.81$\pm$0.02   & -8.34$\pm$0.19    & 0.20    & 0.84    & \citet{2016ApJ...821L..26C} & 306 \\
                & Sbc-Irr Galaxies  & 0.74$\pm$0.01   & -8.095$\pm$0.05   & 0.23    & 0.76    & \citet{2019MNRAS.488.3929C} & 456 \\
                & LSBGs             & 1.04$\pm$0.06   & -10.75$\pm$0.53   & 0.34    &  & \citet{2017ApJ...851...22M} & 56 \\
                & HI-rich LSBGs     & 0.59$\pm$0.02   & -6.15$\pm$0.3     &  &  & \citet{2025PASP..137j4102C} & 277 \\
                \hline
                \multirow{11}{*}{resolved}
                & Full Sample (OLS) & 0.694$\pm$0.0015 & -7.790$\pm$0.012  & 0.351   & 0.635   & This work & 254 (315,239) \\
                & HSBGs (OLS)       & 0.702$\pm$0.0016 & -7.831$\pm$0.012  & 0.345   & 0.652   & This work & 216 (276,681) \\
                & LSBGs (OLS)       & 0.368$\pm$0.0049 & -5.502$\pm$0.037  & 0.319   & 0.356   & This work & 38 (38,558) \\
                & Full Sample (ODR) & 1.149$\pm$0.0019 & -11.295$\pm$0.015 & 0.261   &  & This work & 254 (315,239) \\
                & HSBGs (ODR)       & 1.120$\pm$0.0020 & -11.048$\pm$0.015 & 0.258   &  & This work & 216 (276,681) \\
                & LSBGs (ODR)       & 1.093$\pm$0.0077 & -10.995$\pm$0.058 & 0.269   &  & This work & 38 (38,558) \\
                & ALMaQUEST (ODR)   & 1.19$\pm$0.01    & -11.68$\pm$0.11   & 0.25    &  & \citet{2019ApJ...884L..33L} & 14 \\
                & Sbc-Irr SFAs      & 0.94$\pm$0.08    & -9.88$\pm$0.69    & 0.27    & 0.62    & \citet{2019MNRAS.488.3929C} & 456 \\
                & MaNGA (ODR)       & 1.005$\pm$0.004  & -10.338           & 0.127   &  & \citet{2017ApJ...851L..24H} & 536 \\
                & PHANGS            & 1.05$\pm$0.01    & -10.63            & 0.51    &  & \citet{2021AA...650A.134P}  & 18 (313,227) \\
                & CALIFA            & 1.01$\pm$0.15    & -10.27$\pm$0.22   & 0.16    & 0.63    & \citet{2021MNRAS.503.1615S} & 533 \\
                \hline
        \end{tabular}
        \tablefoot{This table lists the parameters of linear fitting for global and resolved SFMS, $\mathrm{y = ax + b}$, where x is $\mathrm{M_{*}}$ or $\mathrm{\Sigma_{*}}$ and y is SFR or $\mathrm{\Sigma_{SFR}}$. The a, b, $\sigma$, and $\rho$ represent the slope, intercept, root mean square error, and Pearson Correlation Coefficient of linear fitting. The $\sigma$ is defined as the root value of the weighted sum of squared residuals divided by degrees of freedom. The sample sizes in the relation are listed in the last column, the number outside parentheses refers to the number of galaxies, inside to the number of spaxels, used in the relation. For the rSFMS relation, some works provide only the galaxy number and lack the spaxel number.}
\end{table*}

In this section, we check the global and resolved SFMS. For individual galaxies, only ``star forming areas'' (SFAs) are valid in deriving SFMS; SFAs are defined as spaxels located under the pure star formation curve in the BPT diagram, $\rm log([O\Rmnum{3}]/H\beta) = 1.3 + 0.61/(log([N\Rmnum{2}]/H\alpha) - 0.05)$ \citep{2003MNRAS.346.1055K}. The spatial resolution of MaNGA data is 2.5$^{\prime\prime}$, which corresponds to physical resolution between 0.69 $\sim$ 2.76 kpc. The global SFR and $\mathrm{M_{*}}$ are derived from integration of local $\mathrm{\Sigma_{SFR}}$ and $\mathrm{\Sigma_{*}}$ of the SFAs within a galaxy.

Figure \ref{fig:SFMS} (a) shows the global SFMS of our full sample, the red and blue scatters represent HSBGs and LSBGs respectively. We performed linear fitting with the ``ordinary least squares'' method (OLS; dashed gray line) and the ``orthogonal distance regression'' method (ODR; dashed black line), although there is no obvious difference between the best-fit lines of these two methods. The slopes of the best fit are 1.054$\pm$0.032 (OLS method) and 1.188$\pm$0.034 (ODR method). Panel (b) shows the resolved SFMS (rSFMS) of our full sample. Given the large amount of spaxel, we adopted a 2D histogram colored by spaxel number to express the relationship. The contours enclose 20\%, 40\%, 60\%, and 80\% of the data. The solid gray line and solid black line represent the best-fit lines for OLS and ODR method, respectively. Different to the global relation, there is significant variance between the best-fit lines of these two methods. The slopes of the best fit are 0.694$\pm$0.0015 (OLS method) and 1.149$\pm$0.0019 (ODR method). The parameters of linear fitting are listed in Table \ref{tab:SFMS}.

Many studies have also adopted both OLS and ODR fitting methods, and these two approaches yield different best-fit lines \citep{2017ApJ...851L..24H, 2021MNRAS.501.4777E}. We also found that the parameters given by the OLS and ODR methods show significant differences, but the ODR method provides a better fitting result that captures the trend of the data distribution more accurately compared to the OLS method. This is because the ODR method minimizes orthogonal distances rather than vertical distances (OLS method), when the error of the x variable is not negligible. 

We repeated the same analysis for HSBGs and LSBGs, and the results are shown in Figure \ref{fig:SFMS} (c) $\sim$ (f). When compared horizontally, the full sample and two subsamples (HSBGs and LSBGs) show similar slopes between their global and resolved SFMS, which suggests that the SFMS still holds on kiloparsec scales and the global main sequence may be the summation of a more fundamental relation on small scales. When compared vertically, for both global SFMS and rSFMS, the slopes of the best fit of HSBGs and LSBGs are very close to those of the full sample, which suggests that $\mu_{0}$ does not influence the global SFMS and rSFMS. The rSFMS for each $\mathrm{M_{*}-\mu_{0}}$ bin are also provided in Appendix \ref{sec:appendix_rSFMS}. The slopes are consistent throughout most of the bins, with the exception of two bins, which confirms the previous conclusion that the rSFMS is a more universal relation independent with $\mathrm{M_{*}}$ and $\mu_{0}$. The slope deviation of the highest-mass LSBG bin ([$\mathrm{M_{*}}$: 10.0 - 10.5; $\mu_{0}$: 22.0 - 23.0]) may be related to the relatively small number of galaxies and spaxels in that bin. The reason for the slope deviation of the lowest-mass HSBG bin ([$\mathrm{M_{*}}$: 9.0 - 9.5; $\mu_{0}$: 21.0 - 22.0]) remains unclear and requires further investigation.

\begin{table*}
        \caption{Parameters of linear fitting for rSFMS based on the decorrelated subsamples\label{tab:rSFMS_49}}
        \renewcommand{\arraystretch}{1.2}
        \centering
        \begin{tabular}{cccc}
                \hline\hline
                Sample & a\_ave & b\_ave & Number of spaxels \\
                \hline
                Full Sample & 1.149$\pm$0.012 & -11.296$\pm$0.091 & 6363$\sim$6544 \\
                HSBGs       & 1.120$\pm$0.013 & -11.048$\pm$0.099 & 5582$\sim$5754 \\
                LSBGs       & 1.097$\pm$0.102 & -11.023$\pm$0.772 & 772$\sim$811 \\              
                \hline
        \end{tabular}
        \tablefoot{This table lists the linear fitting results for the rSFMS, $\mathrm{y = ax + b}$, derived from the spatially decorrelated subsamples. The columns list the mean slope (a\_ave) and intercept (b\_ave) with their $1\sigma$ standard deviations across the subsamples, as well as the range (minimum$\sim$maximum) of the number of spaxels in each subsample.}
\end{table*}

\section{Discussion} \label{sec:Discussion}
\subsection{The effects of correlated spaxels on rSFMS}
In the process of linear fitting of the rSFMS, all spaxels defined as SFAs are included. However, the spaxel size of the MaNGA data (0.5$^{\prime\prime}$) is smaller than its spatial resolution (2.5$^{\prime\prime}$). This oversampling introduces correlations between adjacent spaxels, which may affect the derived relations. As mentioned in the MaNGA DAP documentation \citep{2019AJ....158..231W}, adjacent spaxels separated by fewer than five to six spaxels are spatially correlated, which is consistent with the subsampling of the 2.5$^{\prime\prime}$-diameter fiber beam into 0.5$^{\prime\prime}$ × 0.5$^{\prime\prime}$ spaxels.

To rigorously test whether this affects our results, we performed a subsampling procedure. For each galaxy, we divided the FoV into contiguous blocks of 7×7 spaxels. From each block, we selected the spaxel located at the same relative position within the block to construct a subsample. This ensures that any two selected spaxels are separated by at least six spaxels, effectively breaking the spatial correlation. By repeating this selection for all 49 possible positions within a block, we obtained 49 independent, spatially decorrelated subsamples for each galaxy. 

By combining the subsamples of the same offset across all galaxies, we obtained 49 subsamples for the full sample, as well as for the HSBGs and LSBGs separately. Using these correlation-free subsamples, we repeated the rSFMS fitting and the results are listed in Table \ref{tab:rSFMS_49}. The resulting slopes and intercepts show only small variations across the subsamples, and their mean values are fully consistent with those derived from all the spaxels. In addition, we also provide the range (minimum$\sim$maximum) of spaxel numbers across the subsamples. Even the LSBGs, which have the fewest spaxels, yield around 800 data points per subsample, confirming that the fitting results are reliable. This test also demonstrates that the spatial correlation between spaxels does not bias our conclusions.

\subsection{The effects of different excitations on SFMS}

\begin{figure}[htbp!]
        \includegraphics[width=\hsize]{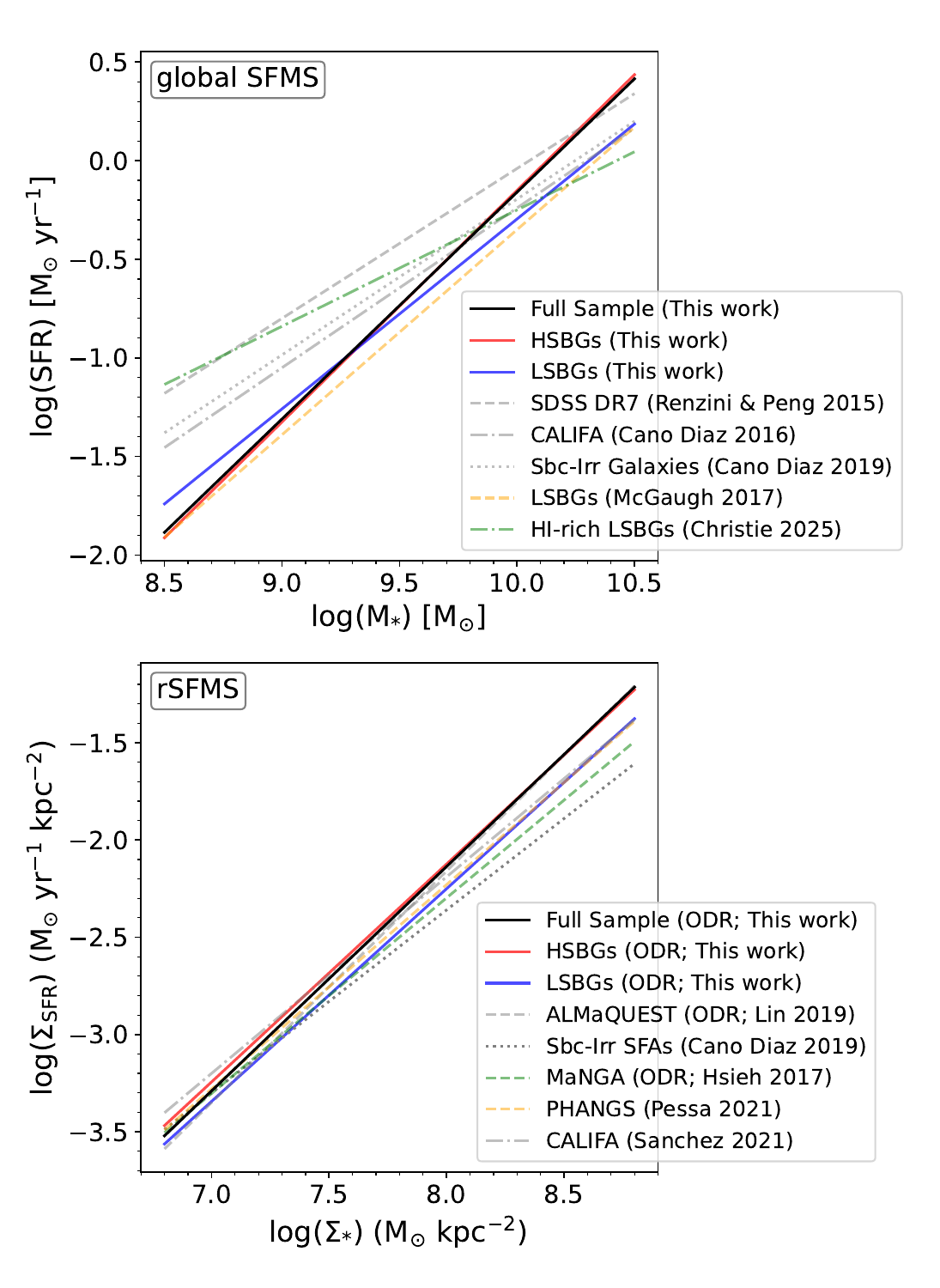}
        \caption{Comparison of the global and resolved SFMS between ours and those in the literature. The black, red, and blue lines represent the SFMS (or rSFMS) of our full sample, HSBGs, and LSBGs, other dashed, dotted, and dash-dotted lines are SFMS (or rSFMS) from references. For rSFMS, our linear fitting result is consistent with that in previous studies, while for global SFMS our slopes are steeper than most of those in the literature. The slope variation between our global SFMS and those from previous studies may arise from different treatments of excitation sources.}
        \label{fig:SFMS_compare}
\end{figure}

The slopes of our SFMS and rSFMS are indistinguishable from unity, and it is reasonable and physically meaningful that the slope of the SFMS is close to 1 \citep{2017ApJ...851...22M}. This is because the stellar mass is the integral of the star formation rate: $\mathrm{M_{*} = \langle SFR \rangle t_{G}}$, where $\mathrm{t_{G}}$ is the age of the galaxy and $\mathrm{\langle SFR \rangle}$ is the average star formation rate throughout the galaxy's life. When we take the logarithm of both sides of the equation, it becomes $\mathrm{log(SFR) = log(M_{*}) - constant}$, providing the SFMS with a slope of 1. 

We list some studies on SFMS in Table \ref{tab:SFMS} and plot them in Figure \ref{fig:SFMS_compare}. For the rSFMS, our linear fitting result is consistent with that in previous studies \citep{2017ApJ...851L..24H, 2019ApJ...884L..33L, 2019MNRAS.488.3929C, 2021AA...650A.134P, 2021MNRAS.503.1615S}, and the slopes are all near to 1. However, for the global one, our slope, which is slightly greater than 1, is steeper than that in some previous studies, which are slightly below 1, though both are close to 1. The slope variation between our global SFMS and those from previous studies \citep{2015ApJ...801L..29R, 2016ApJ...821L..26C, 2019MNRAS.488.3929C, 2025PASP..137j4102C} may arise from different treatments of excitation sources.

Selecting different regions within galaxies dominated by various ionizing sources for constructing SFMS will result in changes to both the position of galaxies on the SFMS and the slope of the SFMS itself. \citet{2018ApJ...854..159P} found that, when the global SFR and $\mathrm{M_{*}}$ are integrated by areas dominated by all ionizing sources in galaxies, the global SFMS will turn over with increasing stellar mass and bulge-to-total ratio (B/T), in turn leading to the flat slope of global SFMS in the literature. \citet{2017ApJ...851...22M} also observed a similar bend of global SFMS at around $\mathrm{M_{*} \approx 10^{10}\ M_{\odot}}$. They attribute this bend to the fact that massive spirals with $\mathrm{M_{*} > 10^{10}\ M_{\odot}}$ contain a larger fraction of galaxies that have already turned off the SFMS, which pulls the overall population away from the main sequence. The same is true for giant low surface brightness galaixes (gLSBGs) in \citet{2023ApJ...959..105D}; the outer, disk-dominated parts of gLSBG tend to follow the SFMS defined by local SFGs, whereas the inner, bulge-dominated parts of gLSBG tend to drag the entire galaxy off the main sequence. By contrast, when only H\Rmnum{2} regions (or SFAs) are considered, there is no noticeable flattening in both global and resolved SFMS \citep{2018ApJ...854..159P}. 

Therefore, when constructing the SFMS, it is insufficient to select only SFGs, as regions excited by non-star-formation sources inevitably exist within galaxies. Consequently, whether using the SED-fitting method (adopted by \citet{2015ApJ...801L..29R, 2025PASP..137j4102C}) or integrating the local $\mathrm{\Sigma_{*}}$ and $\mathrm{\Sigma_{SFR}}$ across the FoV (adopted by \citet{2016ApJ...821L..26C, 2019MNRAS.488.3929C}), the influence of non-star-forming regions on the total SFR and $\mathrm{M_{*}}$ cannot be avoided, unless nearly all regions within a galaxy are excited by star formation, which is more commonly observed in dwarf galaxies \citep{2017ApJ...851...22M} and disk galaxies (this paper). This is the main reason for the slope difference of global SFMS between previous studies and ours. For the rSFMS, we and all of the references listed in Table \ref{tab:SFMS} have distinguished SFAs within galaxies, leading to more consistent slopes among these studies.

\subsection{The limitation of our sample}
\begin{figure*}
        \includegraphics[width=\hsize]{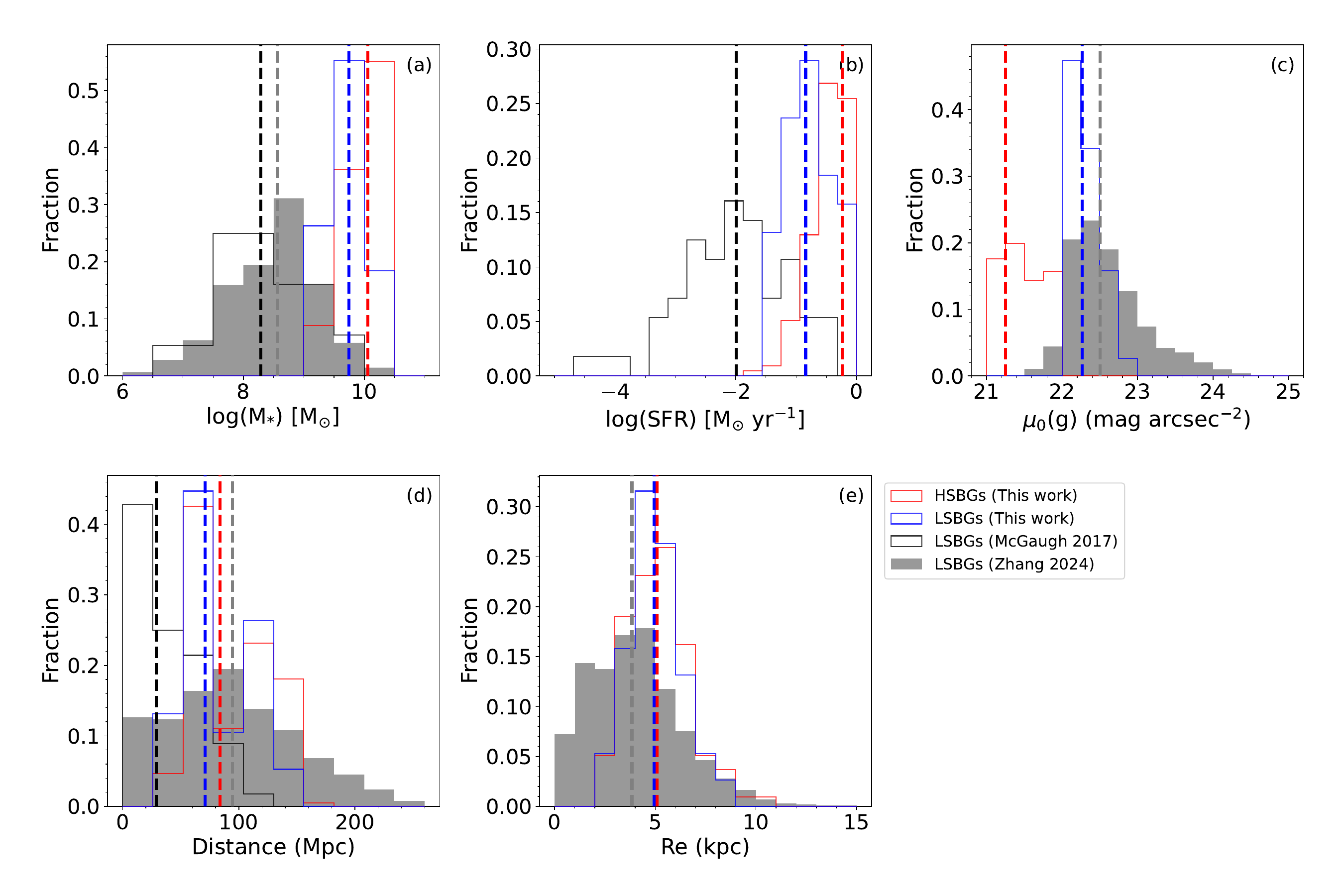}
        \caption{The histograms of stellar mass ($\mathrm{M_{*}}$), central surface brightness ($\mu_{0}$), total SFR, distance, and effective radius ($\mathrm{R_{e}}$) for the following samples: HSBGs from this work (red steps; 216 galxies), LSBGs from this work (blue steps; 38 galaxies), LSBGs from \citet{2017ApJ...851...22M} (black steps; 56 galaxies), and LSBGs from \citet{2024RAA....24a5018Z} (filled gray steps; 1038 galaxies). For each sample, the median value of each parameter is marked by a vertical line using the same color as the histogram. Our sample lacks LSBGs with lower $\mathrm{M_{*}}$, fainter $\mu_{0}$, and lower SFR.}
        \label{fig:Sample_histograms}
\end{figure*}

Figure \ref{fig:Sample_histograms} presents the histograms of $\mathrm{M_{*}}$, SFR, $\mu_{0}$, distance, and $\mathrm{R_{e}}$ for the following samples: HSBGs from this work (red steps; 216 galxies), LSBGs from this work (blue steps; 38 galaxies), LSBGs from \citet{2017ApJ...851...22M} (black steps; 56 galaxies), and LSBGs from \citet{2024RAA....24a5018Z} (filled gray steps; 1038 galaxies). For each sample, the median value of each parameter is marked by a vertical line using the same color as the histogram. 

Compared with LSBG samples in the literature, our sample lacks LSBGs with lower stellar mass ($\mathrm{M_{*} < 10^{9}\ M_{\odot}}$) and fainter central surface brightness ($\mathrm{\mu_{0, g} > 23.0\ mag\ arcsec^{-2}}$). The median values of $\mathrm{M_{*}}$ histogram of LSBGs samples are around $\mathrm{10^{8.29}\ M_{\odot}}$ \citep{2017ApJ...851...22M} and $\mathrm{10^{8.56}\ M_{\odot}}$ \citep{2024RAA....24a5018Z}, and the minimum values of $\mathrm{M_{*}}$ can be $\sim$ $\mathrm{10^{7}\ M_{\odot}}$. The selection strategy of the MaNGA survey contributes significantly to the fact that our LSBG sample does not contain lower stellar mass, because the MaNGA survey provided a near flat distribution in the stellar mass, but their low-mass limit is $\sim$ $\mathrm{10^{9}\ M_{\odot}}$ \citep{2017AJ....154...86W}. Moreover, to achieve a sufficient S/N, galaxies with lower $\mu_{0}$ and SFR tend not to be selected into the final targeting catalog of the MaNGA survey. Studies have shown that massive galaxies generally follow inside-out formation \citep{2016MNRAS.463.2799I, 2020ARA&A..58...99S}, while low-mass galaxies may exhibit very diverse formation modes. Some low-mass galaxies ($\mathrm{<\ 10^{8}\ M_{\odot}}$) have the outside-in growth mode \citep{2020A&A...633A.105C}, some ($\sim$ $\mathrm{10^{8\sim9}\ M_{\odot}}$) show a possible transition from the inside-out toward the outside-in \citep{2013ApJ...764L...1P, 2020A&A...633A.105C}, and some exhibit an episodic star formation history \citep{2016MNRAS.463.2799I}. A sample covering a larger mass range would help in understanding LSBGs comprehensively.

\section{Summary} \label{sec:Summary}
By employing the MaNGA PyMorph photometric VAC (MPP-VAC-DR17) and the MaNGA Deep Learning Morphological VAC (MDLM-VAC-DR17) \citep{2022MNRAS.509.4024D}, we selected a late-type, face-on, disk-dominated, and star-forming galaxy sample from the MaNGA survey, composed of 38 LSBGs and 216 HSBGs, by adopting a central surface brightness threshold $\mu_{0}{\rm (g) = 22.0\ mag\ arcsec^{-2}}$. 
By using products from the pipe3D team \citep{2022ApJS..262...36S}, we obtained the dust-corrected stellar mass density map ($\mathrm{\Sigma_{*}}$) derived from the stellar population analysis results of pipe3D team and the H$\alpha$-based star formation rate density ($\mathrm{\Sigma_{SFR}}$) according to flux maps of emission lines. 
During the calculation of $\mathrm{\Sigma_{SFR}}$, for each galaxy in our sample, we generated a mask to distinguish SFAs, which are defined as spaxels below the pure star formation curve in the BPT diagram \citep{2003MNRAS.346.1055K}, because the method we used to convert H$\alpha$ luminosity into SFR is only valid within H\Rmnum{2} regions. Additionally, the ratio of $\mathrm{\Sigma_{SFR}}$ and $\mathrm{\Sigma_{*}}$ yields the specific star formation rate ($\mathrm{\Sigma_{sSFR}}$). 

We investigated the radial distributions of $\mathrm{\Sigma_{*}}$, $\mathrm{\Sigma_{SFR}}$, and $\mathrm{\Sigma_{sSFR}}$. The radial profiles of $\mathrm{\Sigma_{*}}$ and $\mathrm{\Sigma_{SFR}}$ exhibit a decreasing trend that correlates with decreasing $\mathrm{M_{*}}$ and fainter $\mu_{0}$, while the radial profiles of $\mathrm{\Sigma_{sSFR}}$ decline with increasing $\mathrm{M_{*}}$ and fainter $\mu_{0}$. 
The $\mathrm{\Sigma_{*}}$ and $\mathrm{\Sigma_{SFR}}$ present negative gradients, the slopes of $\mathrm{\Sigma_{*}}$ and $\mathrm{\Sigma_{SFR}}$ gradient both become stepper and the intercepts of those become larger with the increase of $\mathrm{M_{*}}$ and $\mu_{0}$. The LSBGs show very flat $\mathrm{\Sigma_{SFR}}$ gradients ranging from -0.1 to -0.2, which indicates uniform star formation throughout the whole galaxy. The radial profiles of $\mathrm{\Sigma_{sSFR}}$ present positive gradients, reflecting their ``inside-out'' growth. The slopes of the $\mathrm{\Sigma_{sSFR}}$ gradient become flatter with the increase of $\mu_{0}$.

The H\Rmnum{2} region mask is also applied to the previously derived $\mathrm{\Sigma_{*}}$ and $\mathrm{\Sigma_{SFR}}$ maps to obtain the total stellar mass ($\mathrm{M_{*}}$) and SFR within the H\Rmnum{2} regions of a galaxy.
Based on the $\mathrm{\Sigma_{*}}$ and $\mathrm{\Sigma_{SFR}}$ maps, along with their integrated values within the H\Rmnum{2} regions, we explored the global star formation main sequence (SFMS; SFR vs. $\mathrm{M_{*}}$) and resolved SFMS (rSFMS; $\mathrm{\Sigma_{SFR}}$ vs. $\mathrm{\Sigma_{*}}$). Our LSBGs and HSBGs follow the same global and resolved SFMS as the full sample, and $\mu_{0}$ shows no influence on both SFMS and rSFMS. For our full sample and two subsamples (HSBGs and LSBGs), the slopes between the global SFMS and rSFMS are consistent, indicating the SFMS still hold on the kiloparsec scale and that star formation activities are regulated by local events. The way different excitation sources are handled can influence the slope of the SFMS and rSFMS; we therefore recommend that only SFAs are used.

Our sample lacks LSBGs with lower stellar masses ($\mathrm{M_{*} < 10^{9}\ M_{\odot}}$) and fainter central surface brightnesses ($\mathrm{\mu_{0, g} > 23.0\ mag\ arcsec^{-2}}$), which is a limitation imposed by the sample selection of the MaNGA survey. Combining existing and future IFU data to construct a more comprehensive sample will be helpful in understanding the formation and evolution of LSBGs.

\begin{acknowledgements}
We thank the anonymous referee for comments that helped to improve this paper.
This paper is supported by the National Key R\&D Program of China grant (Nos. 2021YFA1600401 and 2021YFA1600400) and the National Natural Science Foundation of China (NSFC, No. 12273052). 
Z.Z. acknowledges the support of the National Natural Science Foundation of China (NSFC, Nos.12503023 and 12588202), the fellowship from the China Postdoctoral Science Foundation (Certificate Number: 2024M763214), and the National Key R\&D Program of China grant (No. 2024YFA1611900). 
We acknowledge the support of the science research grants from the China Manned Space Project (Nos. CMS-CSST-2025-A14, CMS-CSST-2021-A04) and the National Natural Science Foundation of China (NSFC, Nos. 12090041, 12090040, 12073051, 12003043, 11733006, U1931109).
\end{acknowledgements}

\bibliographystyle{aa}
\bibliography{aa60456-26}

\begin{appendix}
\onecolumn      
\section{Some parameters of our LSBGs sample} \label{sec:appendix_FoV}

\begin{table*}[htbp]
        \caption{Parameters of LSBGs sample\label{tab:LSBG_info}}
        \renewcommand{\arraystretch}{1.2}
        \centering
        \begin{tabular}{cccccccc}
                \hline\hline
                Index & MaNGA ID & PLATEIFU & Distance & $\mathrm{log(M_{*})}$ & $\mathrm{log(SFR_{H\alpha})}$ & $\mu_{0}$ & Re\\
                 & & & Mpc & $\mathrm{[M_{\odot}]}$ & $\mathrm{[M_{\odot}\ yr^{-1}]}$ & $\mathrm{mag\ arcsec^{-2}}$ & kpc \\
                \hline
                 1 & 1-322308 & 9870-12701  &  82.63 &  9.21 & -1.38 & 22.30 & 4.21 \\ 
                 2 & 1-592984 & 8335-12704  &  78.98 &  9.24 & -1.41 & 22.17 & 5.69 \\ 
                 3 & 1-60706  & 10516-12705 & 102.01 &  9.28 & -0.86 & 22.51 & 4.22 \\ 
                 4 & 1-148985 & 8999-12704  &  99.78 &  9.31 & -0.85 & 22.38 & 4.48 \\ 
                 5 & 1-230171 & 8942-12705  &  87.87 &  9.32 & -1.30 & 22.64 & 3.44 \\ 
                 6 & 1-419251 & 8322-12703  & 103.35 &  9.33 & -0.88 & 22.60 & 5.50 \\ 
                 7 & 1-77565  & 10842-12701 &  75.08 &  9.33 & -1.28 & 22.26 & 3.83 \\ 
                 8 & 1-457321 & 9879-12705  &  56.92 &  9.40 & -0.84 & 22.06 & 2.69 \\ 
                 9 & 1-144829 & 11958-12705 &  63.03 &  9.42 & -0.96 & 22.51 & 3.45 \\ 
                10 & 1-261390 & 8334-12705  &  80.90 &  9.44 & -1.14 & 22.08 & 2.67 \\ 
                11 & 1-166738 & 8459-12705  &  69.75 &  9.50 & -0.80 & 22.14 & 3.72 \\ 
                12 & 1-1907   & 10846-12704 & 148.31 &  9.52 & -1.04 & 22.07 & 5.03 \\ 
                13 & 1-283638 & 11953-12705 &  61.11 &  9.58 & -0.85 & 22.39 & 3.84 \\ 
                14 & 1-389244 & 8150-12702  &  96.06 &  9.65 & -0.77 & 22.14 & 6.26 \\ 
                15 & 1-42313  & 8096-12701  &  75.82 &  9.69 & -1.21 & 22.39 & 4.55 \\ 
                16 & 1-368219 & 11022-12701 & 100.27 &  9.70 & -1.01 & 22.49 & 4.47 \\ 
                17 & 1-39500  & 8569-12702  &  78.03 &  9.70 & -1.37 & 22.73 & 3.79 \\ 
                18 & 1-164062 & 8941-12701  & 100.00 &  9.72 & -1.12 & 22.13 & 4.70 \\ 
                19 & 1-378182 & 8134-12705  &  77.57 &  9.73 & -0.99 & 22.21 & 4.60 \\ 
                20 & 1-152316 & 11954-12702 &  67.08 &  9.76 & -1.07 & 22.19 & 4.35 \\ 
                21 & 1-29751  & 8654-12703  & 161.83 &  9.79 & -0.44 & 22.51 & 5.50 \\ 
                22 & 1-27363  & 12068-12704 & 162.13 &  9.80 & -0.86 & 22.49 & 6.04 \\ 
                23 & 1-404066 & 11826-12701 &  92.38 &  9.81 & -0.94 & 22.96 & 5.78 \\ 
                24 & 1-403615 & 8321-12702  & 100.42 &  9.81 & -0.71 & 22.30 & 7.04 \\ 
                25 & 1-196804 & 11016-12704 & 116.14 &  9.85 & -0.45 & 22.07 & 5.95 \\ 
                26 & 1-324762 & 9893-12705  & 176.07 &  9.86 & -0.57 & 22.04 & 5.26 \\ 
                27 & 1-578295 & 8569-12704  & 168.85 &  9.86 & -0.34 & 22.30 & 5.26 \\ 
                28 & 1-324693 & 11984-12705 & 162.63 &  9.86 & -0.59 & 22.17 & 4.74 \\ 
                29 & 1-120369 & 8088-12701  & 112.65 &  9.92 & -0.30 & 22.15 & 4.91 \\ 
                30 & 1-38514  & 8085-12705  & 177.69 &  9.94 & -0.43 & 22.39 & 6.11 \\ 
                31 & 1-27404  & 12069-12701 & 164.82 &  9.95 & -0.73 & 22.06 & 5.45 \\ 
                32 & 1-209744 & 9031-12702  & 178.00 & 10.04 & -0.67 & 22.18 & 4.93 \\ 
                33 & 1-294237 & 11981-9101  & 130.98 & 10.06 & -0.28 & 22.04 & 4.94 \\ 
                34 & 1-547120 & 8614-12702  & 105.43 & 10.10 & -0.25 & 22.11 & 6.57 \\ 
                35 & 1-151639 & 11956-12704 & 177.64 & 10.13 & -0.21 & 22.27 & 8.76 \\ 
                36 & 1-289993 & 8086-12701  & 177.88 & 10.26 & -0.49 & 22.44 & 5.50 \\ 
                37 & 1-339295 & 8714-12703  & 193.16 & 10.34 & -0.12 & 22.19 & 7.23 \\ 
                38 & 1-217752 & 9498-12705  & 195.99 & 10.40 & -0.21 & 22.31 & 6.87 \\
                \hline
        \end{tabular}
        \tablefoot{This table lists some of the information of our LSBGs sample, sorted by stellar mass ($\mathrm{M_{*}}$).}
\end{table*}

\newpage
\section{The rSFMS in each $\mathrm{M_{*}}$ - $\mu_{0}$ bins} \label{sec:appendix_rSFMS}
\begin{figure*}[h]
        \includegraphics[width=\hsize]{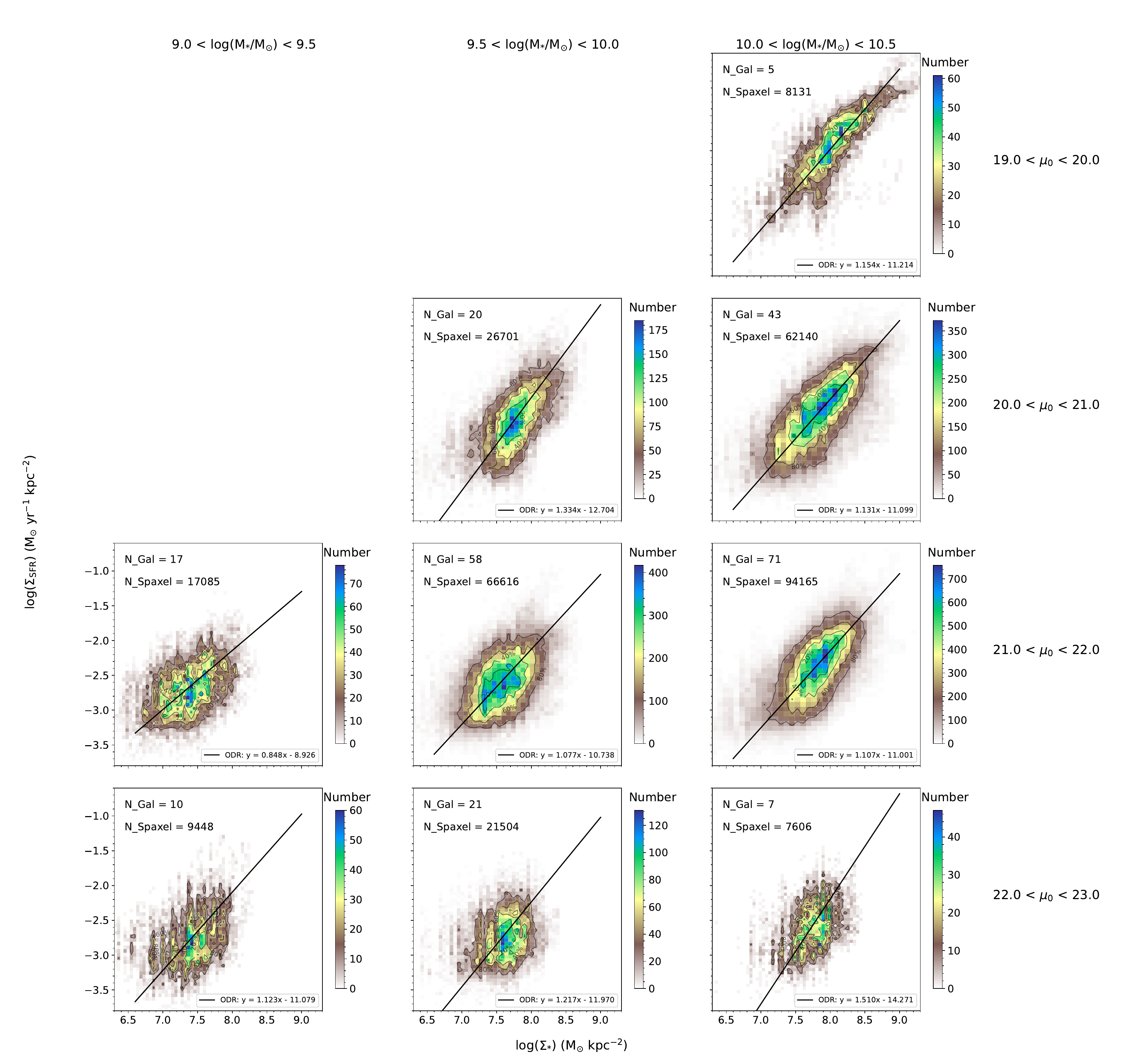}
        \caption{The rSFMS in each $\mathrm{M_{*}}$ - $\mu_{0}$ bins. The linear fitting results are shown in Table \ref{tab:rSFMS_bins}.}
        \label{appendix_fig:rSFMS_bins}
\end{figure*}

\begin{table*}
        \caption{Parameters of linear fitting for rSFMS of each $\mathrm{M_{*}-\mu_{0}}$ bins \label{tab:rSFMS_bins}}
        \renewcommand{\arraystretch}{1.2}
        \centering
        \begin{tabular}{cccccc}
                \hline\hline
                \multicolumn{2}{c}{$\mathrm{M_{*}-\mu_{0}}$ Bins} & a & b & $\sigma$ & Number of \\
                & & & & & galaxies (spaxels)\\  
                \hline
                \multirow{2}{*}{9.0 $\sim$ 9.5} & 21 $\sim$ 22 & 0.848$\pm$0.008 & -8.926$\pm$0.060 & 0.266 & 17 (17,085) \\
                & 22 $\sim$ 23 & 1.123$\pm$0.015 & -11.079$\pm$0.114 & 0.281 & 10 (9,448) \\
                \hline
                \multirow{3}{*}{9.5 $\sim$ 10.0} & 20 $\sim$ 21 & 1.334$\pm$0.008 & -12.704$\pm$0.062 & 0.240 & 20 (26,701) \\
                & 21 $\sim$ 22 & 1.077$\pm$0.005 & -10.738$\pm$0.037 & 0.269 & 58 (66,616) \\        
                & 22 $\sim$ 23 & 1.217$\pm$0.013 & -11.970$\pm$0.097 & 0.266 & 21 (21,504) \\
                \hline
                \multirow{4}{*}{10.0 $\sim$ 10.5} & 19 $\sim$ 20 & 1.154$\pm$0.008 & -11.214$\pm$0.065 & 0.200 & 5 (8,131) \\
                & 20 $\sim$ 21 & 1.131$\pm$0.004 & -11.099$\pm$0.031 & 0.256 & 43 (62,140) \\        
                & 21 $\sim$ 22 & 1.107$\pm$0.003 & -11.001$\pm$0.026 & 0.252 & 71 (94,165) \\        
                & 22 $\sim$ 23 & 1.510$\pm$0.023 & -14.271$\pm$0.178 & 0.230 & 7 (7,606) \\
                \hline  
        \end{tabular}
        \tablefoot{This table lists the parameters of linear fitting for resolved SFMS of each $\mathrm{M_{*}-\mu_{0}}$ Bins, $\mathrm{y = ax + b}$, where x is $\mathrm{\Sigma_{*}}$ and y is $\mathrm{\Sigma_{SFR}}$. The a, b, and $\sigma$ represent the slope, intercept, and root mean square error of linear fitting. The sample sizes in the relation are listed in the last column, the number outside parentheses refers to the number of galaxies, inside to the number of spaxels, used in the relation.}
\end{table*}

\end{appendix}

\end{document}